\documentclass[runningheads]{llncs}

\usepackage{graphicx}
\graphicspath{{./figures/}}

\usepackage{array}
\newcolumntype{C}{>{\centering\arraybackslash}p{2cm}}

\usepackage[hyphens]{url}
\usepackage{subfig}

\usepackage{paralist}

\usepackage{float}

\usepackage{hyperref}

\begin{document}

\title{Keep It Unbiased: A Comparison Between Estimation of Distribution Algorithms and Deep Learning for Human Interaction-Free Side-Channel Analysis}
\titlerunning{A Comparison Between EDAs and DL for Human Interaction-Free SCA}

\author{Unai Rioja\inst{1,2} \and
Lejla Batina\inst{1} \and
Igor Armendariz\inst{2} \and
Jose Luis Flores\inst{2}}
\authorrunning{Rioja et al.}
%
\institute{Digital Security Group, Radboud University, Nijmegen, The Netherlands \email{\{unai.riojasabando,lejla.batina\}@ru.nl} \and
Ikerlan Technological Research Centre, Arrasate-Mondragón, Gipuzkoa, Spain \email{\{urioja,iarmendariz,jlflores\}@ikerlan.es}}

\maketitle

\begin{abstract}
Evaluating side-channel analysis (SCA) security is a complex process, involving applying several techniques whose success depends on human engineering. Therefore, it is crucial to avoid a false sense of confidence provided by non-optimal (failing) attacks.
Different alternatives have emerged lately trying to mitigate human dependency, among which deep learning (DL) attacks are the most studied today. DL promise to simplify the procedure by e.g. evading the need for point of interest selection or the capability of bypassing noise and desynchronization, among other shortcuts. However, including DL in the equation comes at a price, since working with neural networks is not straightforward in this context.
Recently, an alternative has appeared with the potential to mitigate this dependence without adding extra complexity: Estimation of Distribution Algorithm-based SCA. In this paper, we compare these two relevant methods, supporting our findings by experiments on various datasets.
\end{abstract}


\section{Introduction}

These days we are surrounded by IoT devices that handle sensitive information, not only in industrial applications but also in our daily lives. This requires from designers and product developers to use cryptography to protect embedded devices, but cryptography is only one of the components ensuring the security of systems. As Kocher \textit{et al.} demonstrated in 1999~\cite{kocher1999dpa}, the security of a device depends not only on the mathematical characteristics of its cryptographic operations but also on its physical implementation. In other words, the physical nature of these devices can be exploited in order to break their security in several ways. Whereas some approaches are passive and rely on simply observing certain physical properties to retrieve information (Side-channel analysis, SCA), other procedures try to stress the system to alter its natural behaviour (Fault Injection, FI). Physical attacks are truly powerful, as they can bypass hardware and software security measures that the manufacturer has included in the design.

The problem is that the inclusion and validation of countermeasures against this kind of attacks is not simple, especially in the SCA case. Current certification schemes require attacking the device under test (DUT) with a battery of known SCA attacks to prove its security~\cite{Melissa2020Systematic}. Unfortunately, this approach is prohibitive in terms of time and resources. The ever-growing number of attack techniques, which in turn involve knowledge of very diverse subjects (e.g., signal processing, electronic design, cryptography, statistics, machine learning, etc.), make it difficult to master and correctly apply all of them. Moreover, the outcome of such attacks depends to a large extent on the experience of the person performing them. Therefore, a process so dependent on human interaction can lead to a biased result if the tests are not properly executed.

One of the most prominent techniques in SCA research today is Deep Learning (DL)-based SCA~\cite{kim2018noise,Maghrebi2016BreakingCI,Cagli2017CNNagainstCM,prouff2018ascad,Rijsdijk2021Reinforcement,Wouters2020Revisiting}, part of the so-called profiling attacks (the strongest SCA technique nowadays). This method aims to overcome some drawbacks of classical SCA, such as the need for pre-processing or point of interest (POI) selection, promising a relaxation of the evaluator interaction. Conversely, working with DL is complex, especially considering the large number of possible architectures and hyper-parameters to adjust. Besides, although some attempts have been made~\cite{kim2018noise,Zaid2019Methodology,Wouters2020Revisiting}, the SCA community has not yet agreed on how models should be constructed and evaluated. In addition, there is no generalized solution: when attacking new datasets, devices or cryptographic implementations, sometimes it is necessary to completely readjust the neural network. Consequently, while some of the more common difficulties are alleviated, new DL-related complications emerge.

Recently, an option that promises to lessen this human dependency without increasing the overall complexity has emerged: Estimation of Distribution Algorithm-based Profiling Attacks (EDA-based PAs)~\cite{Rioja2021AutoTune,Rioja2021Elimination}. This concept comprises applying randomized optimization heuristics to the POI selection issue, allowing an automated boosting of the whole attack (POI selection, leakage profiling and key recovery). This method can produce state-of-the-art results straightforwardly, even in noisy environments~\cite{Rioja2021Elimination}. Although they provide a simple alternative to DL, to the best of our knowledge, both approaches have never been directly compared. Therefore, in this paper, a comparison between the two methods is driven.\\

Thus, the contributions of this work are the following:
\begin{itemize}
    \item A comparative analysis of both approaches has been driven, in terms of complexity and performance, to highlight the strengths and weaknesses of each method.
    \item Our findings are based on repeatable experimental evidence. Some experiments have been carried out especially for this paper, while for others we have relied on related works to provide a comprehensive and impartial comparison.
    \item We also study the performance of several SCA-oriented DL architectures in unfavourable conditions (no mask leakage during the targeted window) to determine whether the same conclusions as in~\cite{Rioja2021Elimination} using EDAs hold.
\end{itemize}

The paper is organized as follows: In Sect.~\ref{sec:Background} we summarize relevant background on profiling attacks. We describes the related work that is closest to ours in Sect.~\ref{sec:RelWork}. We present the comparison of the two techniques (EDAs and DL-based) in Sect.~\ref{sec:Comp}. In Sect.~\ref{sec:Discussion} we discuss the comparison and elaborate on our findings. Sect.~\ref{sec:Concl} concludes the paper.

\section{Background on profiling attacks}
\label{sec:Background}

\begin{figure}[!htb]
	\centering
	\resizebox{1\textwidth}{!}{
	\includegraphics{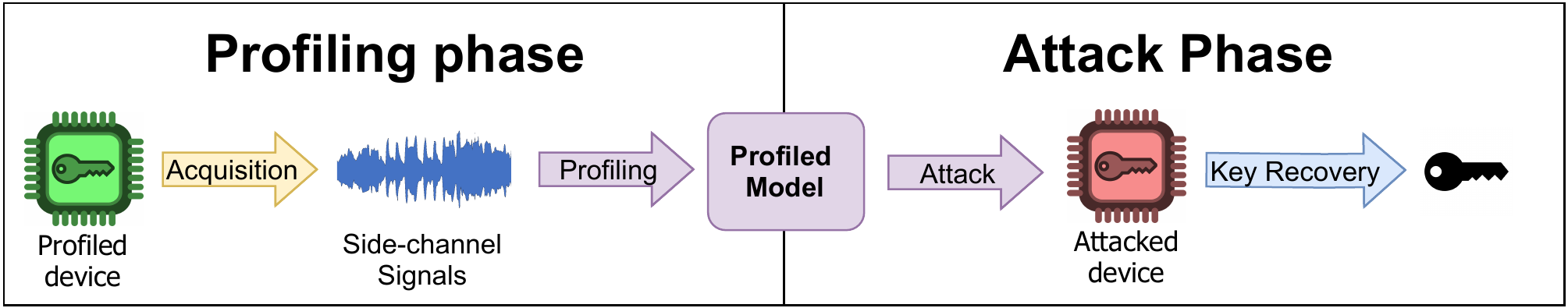}}
	\caption{\label{fig:PAs} Scheme of a generic Profiling Attack}
\end{figure}

Profiling Attacks (PAs) have become the prevailing type of SCA attack in recent years~\cite{Choudary2018efficient,prouff2018ascad,Zotkin2018DeepLV,Zaid2019Methodology,Rioja2021Elimination}. The original idea comprises generating a model of the power consumption of a device to be deployed for the secret key recovery. These attacks involve two phases (See Fig.~\ref{fig:PAs}): a phase in which the model is generated using a typically big number of traces (profiling phase), and the phase in which this model is compared with the actual power consumption of the device to recover sensitive information (attack phase).

Depending on the method used to generate the model, there exist different types of profiling attacks. Whereas the first publications employed gaussian classification (Template Attacks, TAs)~\cite{Chari2002template}, other approaches use linear regression (Stochastic Models approach~\cite{Schindler2005stochastic}) or even Machine Learning (ML) techniques such as Support Vector Machine (SVM)~\cite{Hospodar2011svm,Lerman2011svm,Heuser2012svm}, Random Forest (RF) \cite{Lerman2015rf} or recently introduced Deep Learning (DL) ~\cite{Maghrebi2016BreakingCI,Cagli2017CNNagainstCM,picek2018ontheperformance}. In this paper we focus on TAs and DL as they are the most widely used options in practice~\cite{Zotkin2018DeepLV,Lerman2018Template}. 

\subsection{Template Attacks}

Template attacks are the first type of profiling attacks introduced and they involve building a multivariate model of the probability distribution of the leakage. The practice is to use an extensive set of power traces taken from the DUT when it is manipulating some intermediate value $ v = f(p,k) $. As long as $v$ is related to a known variable (usually plaintext $p$ or ciphertext $c$) and the secret key $k$, guessing $v$ allows the attacker to disclose the secret key. 

First, in the \textit{profiling phase}, the attacker employs a set of $ n_p $ profiling traces ($ \mathbf{T}_{p}$) to build a Gaussian multivariate model, for each possible $v$, creating the so-called \textit{templates} (pairs of mean vectors and a covariance matrices $(\mathbf{m},\mathbf{C})$). Note that, as each $v$ depend on $d$ and $k$, each key hypothesis suggests a template. Finally, a \textit{discriminant score} $ D\left(k \mid \mathbf{t}_{i} \right) $ is computed for each key hypothesis $ k_{j} $ and all key hypothesis are graded in decreasing order of probability, creating a key guessing vector. In SCA, it is common to work with a metric called Guessing Entropy ($ge$)~\cite{Standaert2009UnifiedFramework}, which is the average rank of the correct candidate $k^{*}$ after several attacks. 

In practice, TAs pose several limitations such as complexity issues or the need for dimensionality reduction~\cite{Choudary2018efficient}. The latter is often fixed by choosing just a few time samples of the power traces (POIs selection \cite{Rechberger2005Practical}), or employing a more complex technique like Principal Component Analysis (PCA)~\cite{Archambeau2006subspaces,Standaert2008Using} or Fisher’s Linear Discriminant Analysis (LDA)~\cite{Jhonson2009,LDA2}). Note that, with EDA-based PA, the POI selection is performed automatically by the algorithm~\cite{Rioja2021AutoTune}. 

\subsection{DL-based SCA}
In recent times, multiple papers related to DL-based SCA have been published. The approach is the same as for template attacks, but the model building and classification are performed using neural networks. Most of these works rely on two typical deep learning architectures: Multilayer Perceptron (MLP) and Convolutional Neural Networks (CNN). The early architecture used in DL-SCA for its simplicity was MLP. The first proposal was applying regression to characterize leakage \cite{yang2011backprop} but the approach rapidly evolved to use MLP as a classifier for intermediate values of the traces~\cite{Maghrebi2016BreakingCI,Martinasek2014comparison,martinasek2015profiling}. After this, CNNs also began to be used because its spatial invariance property provides robustness against data distortions like environmental noise, desynchronization and countermeasures~\cite{martinasek2015profiling,Maghrebi2016BreakingCI,Cagli2017CNNagainstCM,hettwer2018profiled,prouff2018ascad,picek2018ontheperformance}. Other studies have examined the performance of various PAs~\cite{Zotkin2018DeepLV,picek2018ontheperformance,prouff2018ascad,Zaid2019Methodology}.

As stated above, one of the major drawbacks of classical SCA is the need for pre-processing and POI selection, as this part is strongly dependent on human engineering. DL-SCA claims to overcome those struggles, since the features are selected automatically from traces by the neural network. In any case, note that although this part of the problem may be mitigated, working with neural networks is a complex process that also requires human interaction for architecture selection, tuning and training of the neural networks to operate correctly.

\subsection{Masking protected implementations}

SCA countermeasures try to obfuscate the dependency between the power consumption of the DUT and the intermediate values of the implemented cryptographic algorithm. One of the most popular ones is masking~\cite{BlueBook}. Masking (also known as secret sharing) comprise concealing each intermediate value $ v $ with a random value $ m $ (mask), which is different for each execution and unknown by the attacker, such as $ v_{m} = v * m $. If correctly implemented, this ensures pairwise independence between masked values, unmasked values and the masks. Consequently, a classical (first-order) Differential Power Analysis (DPA~\cite{kocher1999dpa}) attack would fail. 

Although theoretically sound, implementing masking incorrectly can be fatal for a device’s security. Close manipulation of the shares can provoke unintended interactions between values in the microcontroller, principally caused by transitional effects~\cite{BalaschJosep2014OtCo} and glitches~\cite{Zhimin2009glitches,Mangard2005glitches}. These phenomena can halve the security order, making the first order masking (with two shares or mask and masked intermediate value) vulnerable even to first-order attacks~\cite{BalaschJosep2014OtCo}. This is important since this is the most widely used scheme in practice, because of the complexities involved in higher-order masking~\cite{Shelton2021Rosita}.

\section{Related Work}
\label{sec:RelWork}

\subsection{EDA-based SCA} 

The usage of Estimation of Distribution Algorithms (EDAs) in the SCA context was introduced in~\cite{Rioja2021AutoTune} as an alternative to adjust the POI selection and perform template attacks over embedded devices in an automated way. EDAs are a class of population based optimization heuristics that explore potential results by forming explicit probabilistic models of promising candidates. The approach is to seek in the space shaped for all groupings of POIs for the best ones. Instead of an exhaustive enumeration of all possible combinations, this method performs a search based on a quality measure combined with EDAs. 

In a nutshell, in the first place, an initial population of $R$ individuals (POI selection candidates) is generated. This population can be generated at random or according to some criterion (i.e., assigning a higher probability to samples that present a stronger correlation with the processed intermediate values~\cite{Rioja2021AutoTune}). After this, $R$ attacks are performed with the $R$ candidates and the candidates are rated according to their performance. Then, the best $N$ candidates are chosen ($N<R$) and the probability distribution $p(\mathbf{x})$ of potential candidates is estimated from them. The process is repeated until a stop condition is reached (see~\cite{Rioja2021AutoTune} for more details).

Although they present several advantages against DL, the authors of~\cite{Rioja2021AutoTune,Rioja2021Elimination} have contrasted it with other classical (manual) POI selection techniques but have never compared it directly with DL-based attacks. Thus, in this paper, we compare both techniques on various datasets, in terms of performance and complexity.

\subsection{Profiling Attacks on Masking}

On the one hand, although PAs are a derivative of DPA attacks, many papers claim that (first-order) DL-based SCA attacks can deal with masking countermeasures~\cite{gilmore2015neural,Maghrebi2016BreakingCI,prouff2018ascad,timon2019nonprofileddl}.  The claim is that, in these attacks, the network is trained without mask information (the traces are labelled with the unmasked intermediate value $ v $). Despite this, as deep neural networks can implement highly complex functions, they might be able to guess the correct key without needing this information, and consequently, bypassing masking~\cite{Perin2018LoweringTB}. 

This is remarkable since, with classical TAs, when the mask value is unknown to the attacker during the profiling step, the leakages associated with a key follow a multimodal distribution. This leads to assumption errors whether the adversary exploits Gaussian template attacks, as confirmed in~\cite{Lerman2018Template}. For this reason, some previous works apply TAs combined with second-order techniques and template-based DPA attacks with extra calculation considering the masks to succeed~\cite{Oswald2007template}. When the attacker has such strong capabilities (i.e., knows the key and the masks during profiling), it is considered a worst-case scenario~\cite{Melissa2020Systematic,Bronchain2021Breaking,Bronchain2021Give}.

Nonetheless, as mentioned above, if masking is not properly implemented the implementation can be vulnerable to first-order attacks: It is important to ensure that the two parts of the same secret (mask and masked intermediate value) are not handled too closely~\cite{Shelton2021Rosita}. For instance, authors in~\cite{Zotkin2018DeepLV} claim that when the mask leakage is included in the observation time window, (first-order) TAs can relate the dependence between the mask and the masked variable leakage. Many other related works perform TAs over masked implementation without mask information~\cite{kim2018noise,prouff2018ascad,Rioja2021AutoTune,Rioja2021Elimination}. This allows for a more ``realistic'' (real-world) attack.

To the best of our knowledge, most of the results obtained with DL perform the attack in a weak setting (mask leakage in the attacked window and/or unintended interactions). This is partly because most of the results on DL-based SCAs are based on the ASCAD~\cite{prouff2018ascad} dataset, in which there exist mask leakage in the targeted window, indicating possible unintended interactions. Thus, it is not clear that state-of-the-art CNNs have any advantage over TAs in these conditions, since both can circumvent masking. It is also unclear whether the attack works because of the presence of mask leakage or unintended interactions. 

In any case, the authors of~\cite{Rioja2021Elimination}, published a dataset containing traces from masking implementations with and without mask leakage, but they only use EDA-based attacks. In this paper, we perform DL-based attacks on that dataset trying to determine whether DL-based attacks can bypass masking on both conditions or, on the contrary, current CNNs do not have any advantage in this scenario\footnote[1]{Note that a second-order attack (combining the leakage of two bytes of the key at a time to remove the mask) is feasible in both situations~\cite{Rioja2021Elimination}.}.

\section{Experimental Comparison}
\label{sec:Comp}

In this section, we compare the performance of EDAs and DL-based attacks on two open datasets: ASCAD~\cite{prouff2018ascad} and AES\_RA~\cite{Rioja2021Elimination}.

\subsection{ASCAD Random Keys}

ASCAD \cite{prouff2018ascad} was the early open database for DL-SCA, and has become a standard for experimentation with DL in SCA, with many papers using it appearing every year~\cite{Zaid2019Methodology,Wouters2020Revisiting,Rioja2021Elimination,Rijsdijk2021Reinforcement,masure2019comprehensive,Wu2020OnPre,Hoang2020Plaintext,Zhang2020Anovel}. The DUT in this data set is an 8-Bit AVR microcontroller (ATmega8515), and includes EM emanation traces of the device implementing a masked AES-128 cipher \cite{AES,BlueBook}. The dataset is divided into two parts, fixed key and variable key. Although many works employ the fixed key version because it is an easier problem, for this comparison, we focus on the variable key subset because it is a more challenging and realistic experiment. The traces represent a window of 1400 relevant samples per trace, corresponding to the third byte of the first round masked S-Box operation. As the sensitive intermediate value it is common to use an S-box output: $Y^{(i)}(k^{*})=Sbox[P_{3}^{(i)}\oplus k^{*}]$. For a deeper explanation of the ASCAD dataset, we refer to \cite{prouff2018ascad}.

For this dataset there exist published papers using both DL~\cite{Perin2020strength,Wu2020ICY,Rijsdijk2021Reinforcement} and EDAs~\cite{Rioja2021Elimination}. Thus, we have focused on these works without performing additional experiments, allowing an objective and unbiased view. Table~\ref{tab:DLascad} summarizes the best published attacks against ASCAD, in terms of trainable parameters and $ge$. As usual in the field, for $ge$ we utilize $\bar{Q}_{t_{ge}}$, or the average number of traces needed to obtain a $ge$ of 0.

\begin{table}[!h]
    \centering
    \resizebox{1\textwidth}{!}{
    \begin{tabular}{|c|c|c|c|c|c|}
        \hline
        \textbf{ASCAD Random} & \textbf{\cite{Perin2020strength} Best CNN } & \textbf{\cite{Wu2020ICY} Best CNN }& \textbf{\cite{Rijsdijk2021Reinforcement} Best CNN} & \textbf{\cite{Rijsdijk2021Reinforcement} Best CNN (RS)} & \textbf{\cite{Rioja2021Elimination} Best EDA} \\
        \hline
        \textbf{Trainable Param.} & \textbf{N/A} & 2 076 744 & 70 492 & 3 298 & \textbf{1 400} \\
        \hline
        $\mathbf{\bar{Q}_{t_{ge}}}$ & \textbf{105} & 1 568 & 490 & 1 018 & \textbf{150} \\
        \hline
    \end{tabular}}
    \caption{Top results with PAs on ASCAD random}
    \label{tab:DLascad}
\end{table}

Overall, we can see how despite being easier to execute, EDA-based PAs are able to provide better results (smaller $\bar{Q}_{t_{ge}}$) of than most CNNs on ASCAD (random key), being a more practical and simpler option for evaluators. 

In terms of trainable parameters, the EDA-Based TA is also less demanding (one per time sample). Note that apart from the number of trainable parameters, with DL there is a huge number of possible architectures and hence hyper-parameters to select and adjust. Conversely, the EDA-based approach has a far narrower number of parameters to tune (\# iterations, population size and evaluation function). Finally, the complexity of the ``training’’ itself is again lighter in an EDA-based attack: The time complexity of the EDA (UMDA) is $O(n)$~\cite{Rioja2021AutoTune} whereas the time complexity of a backpropagation algorithm is much larger ($O(n^5)$ according to~\cite{kasperfred}).

\subsection{AES\_RA}

AES\_RA was introduced in~\cite{Rioja2021Elimination} and contains traces from two distinct embedded systems which use microcontrollers from the same family (Piñata training board from Riscure~\cite{PinataDatasheet} and STM32F411E-DISCO development board~\cite{ucDISCO}). With each device, authors acquire traces from three AES implementations: an unprotected software AES implementation and two masking schemes, resulting in six different setups. Thus, this dataset is divided into two parts: power consumption traces from the Piñata board and capacitor EM power traces from the STM32F411E-Discovery Board. Whereas power traces correspond to clean (noise-free) measurements, EM traces represent a more challenging problem (due to noise). Regarding the implementation, the authors show how masked scheme 1 (MS1 from now on) is completely weak against (first-order) PAs, as the mask leaks in the same time window as the masked intermediate value (SBox output). Conversely, masked scheme 2 (MS2) contains only intermediate value leakage in the targeted time window, and hence, this implementation is robust against (first-order) PAs. 
For more details, we refer to the original paper~\cite{Rioja2021Elimination}.

In this work, we perform DL-based experiments on the dataset, confirming that the same conclusions are obtained as in~\cite{Rioja2021Elimination} with EDAs on the different masking schemes and configurations. 

\vspace{1cm}
\subsubsection{Experimental configuration}

We have performed several DL-based attacks on each protected implementation using some (pre-defined) CNN architectures. These architectures were not created for this specific dataset, and hence, good outcomes are not guaranteed. Nevertheless, all were designed for ASCAD, which represents a problem analogous to MS1 (mask leakage on the targeted window). Thus, some CNNs shall succeed, especially considering that MS1 is particularly weak on Piñata (secret key can be recovered with about 5 traces using TAs). This also gives us a clue about how difficult it is to attack a similar dataset with a pre-defined architecture. Regarding trainable parameters, we have selected complex and simple CNNs in order to have a wide range of results (see Table~\ref{tab:ParametersCNNs}).

More precisely, we have used the architecture for ASCAD random key suggested in the original paper~\cite{prouff2018ascad}, four CNNs from~\cite{Rijsdijk2021Reinforcement}, the architecture from~\cite{Zaid2019Methodology} for ASCAD fixed key and an improved version introduced in~\cite{Paguada2020Controlling}:

\begin{itemize}
    \item CNN1: From the original paper~\cite{prouff2018ascad}.
    \item CNN2: From~\cite{Zaid2019Methodology}, for ASCAD Fixed.
    \item CNN3: From~\cite{Paguada2020Controlling}, for ASCAD Fixed.
    \item CNN4: From~\cite{Rijsdijk2021Reinforcement}, ASCAD Random (HW)
    \item CNN5: From~\cite{Rijsdijk2021Reinforcement}, ASCAD Random RS (HW)
    \item CNN6: From~\cite{Rijsdijk2021Reinforcement}, ASCAD Fixed (HW)
    \item CNN7: From~\cite{Rijsdijk2021Reinforcement}, ASCAD Fixed RS (HW)
\end{itemize}

For all of them, we have used the hyper-parameters as in the original articles, except for the number of epochs. We have repeated the experiments with 50 and 75 epochs to ensure that poor results are not obtained because of underfitting/overfitting.

\vspace{1cm}
\subsubsection{Experimental results}

\begin{figure}[H]
	\centering
    \subfloat{\includegraphics[width=0.49\textwidth]{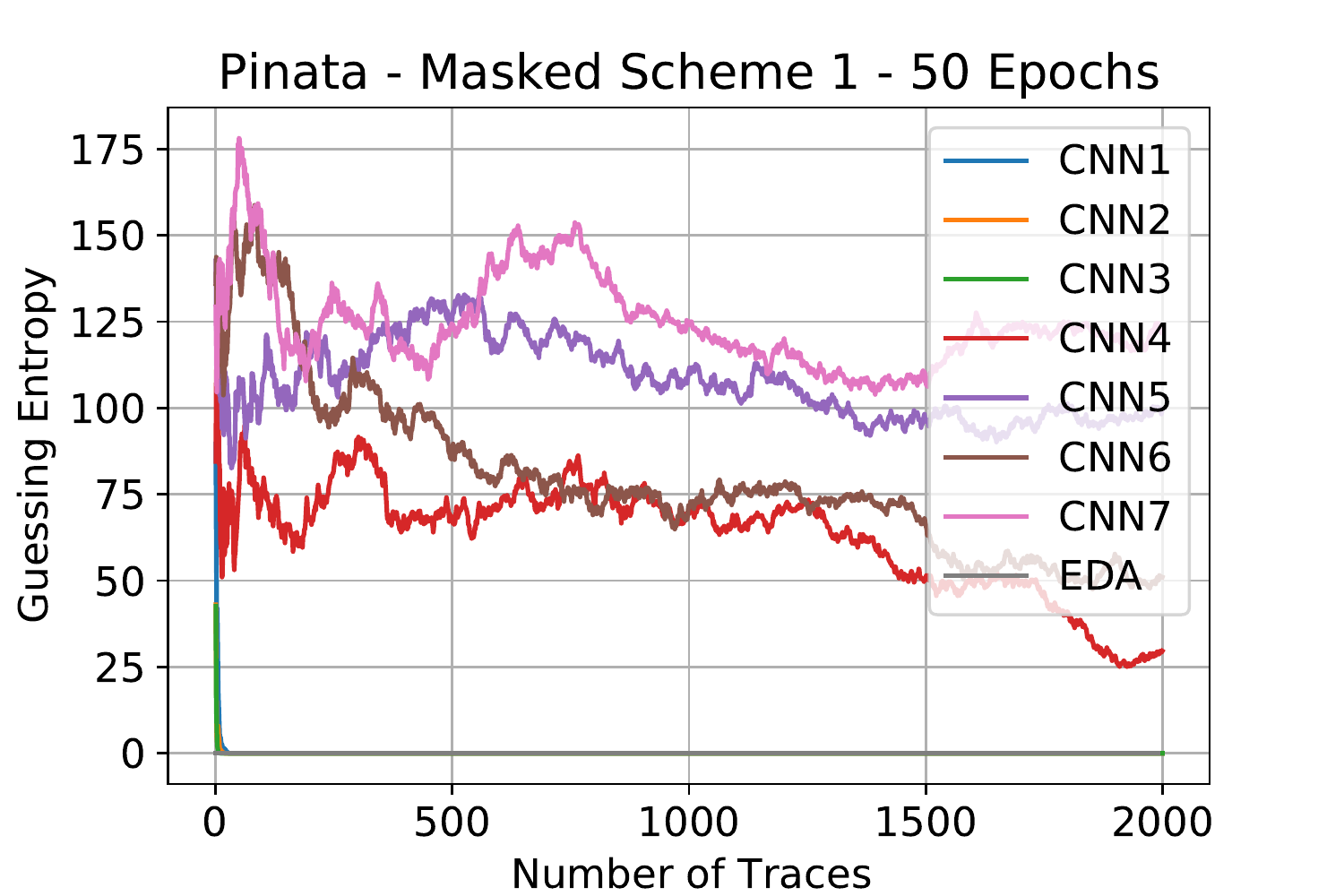}} 
    \subfloat{\includegraphics[width=0.49\textwidth]{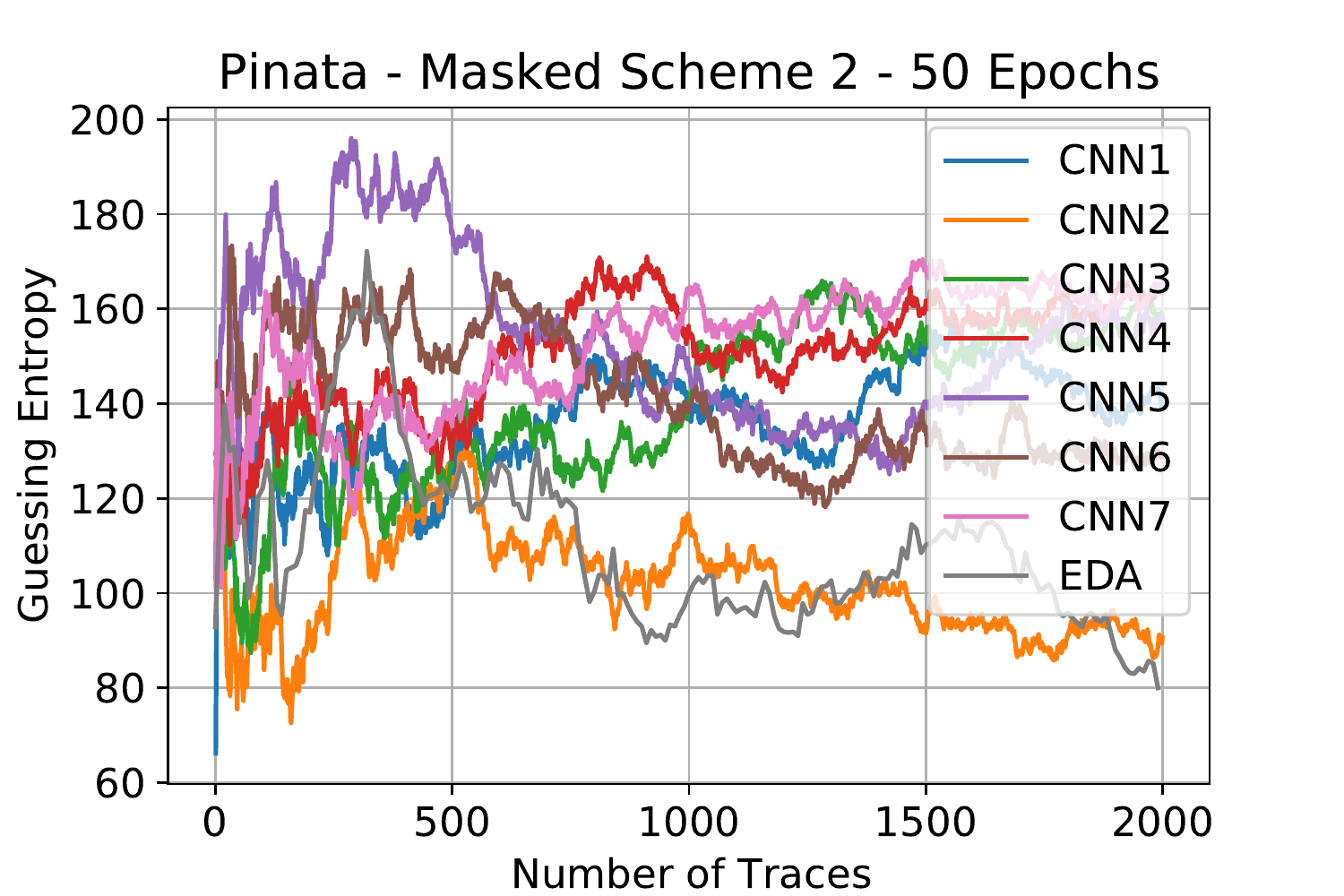}} \\
    \subfloat{\includegraphics[width=0.49\textwidth]{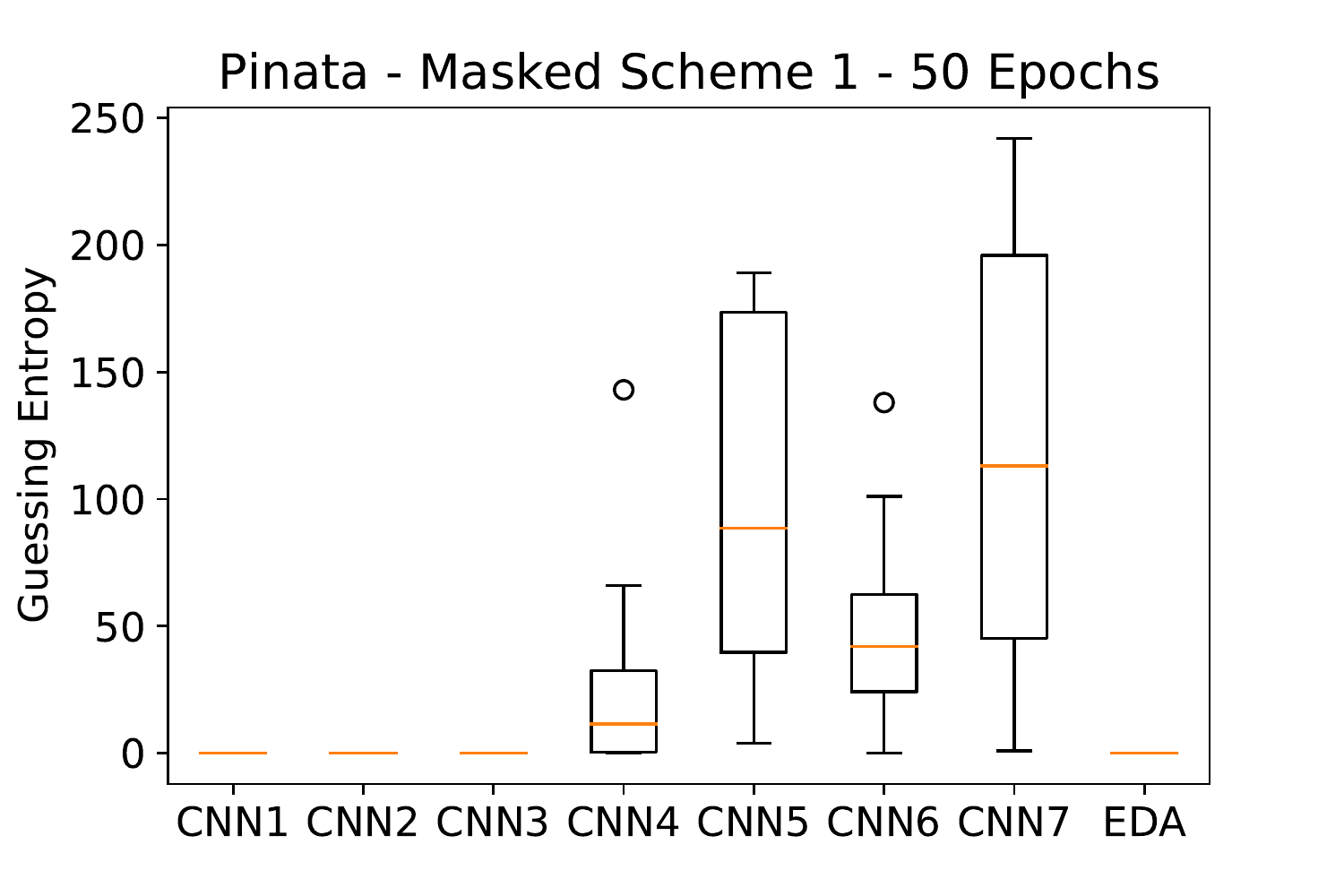}} 
    \subfloat{\includegraphics[width=0.49\textwidth]{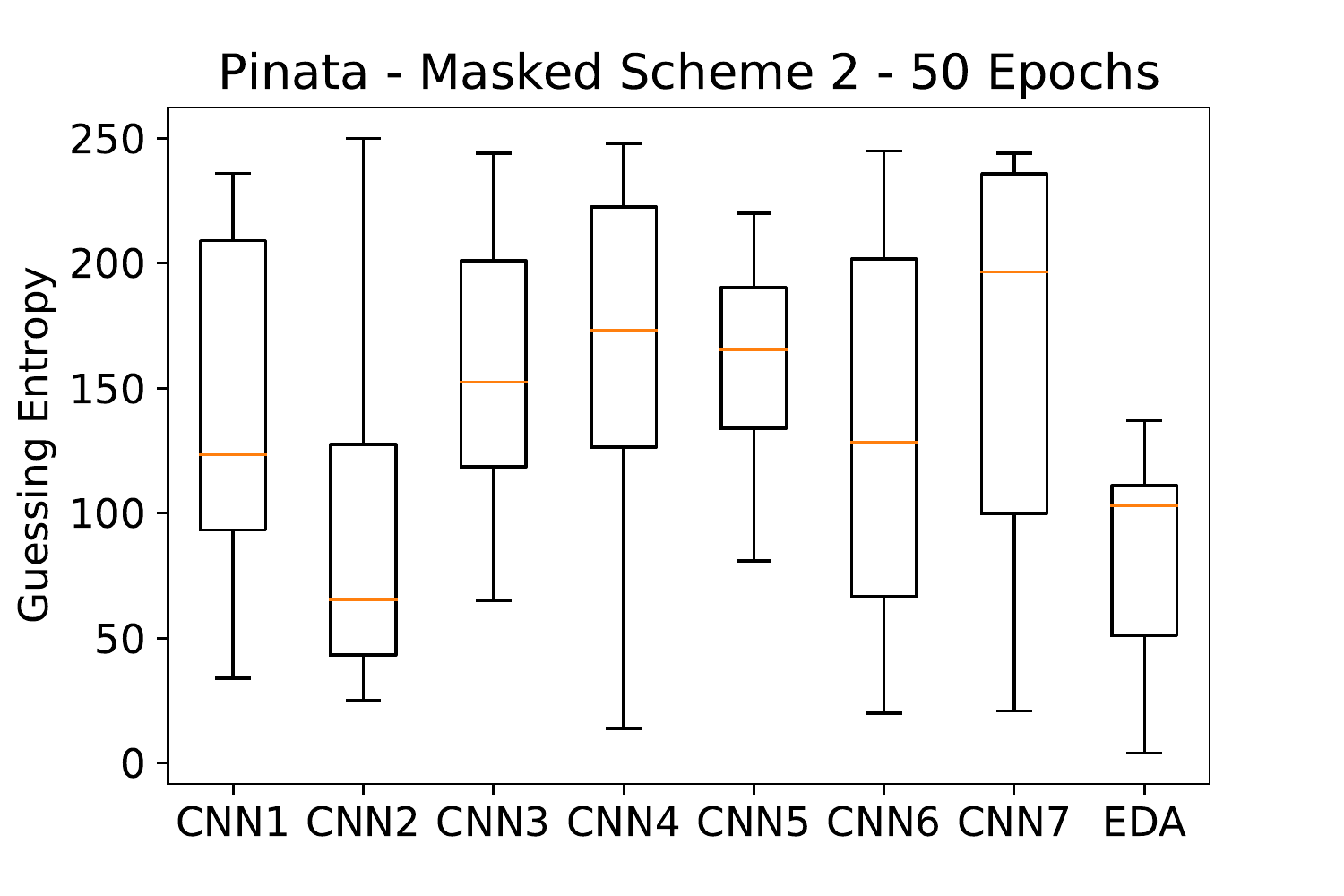}} 
    \caption{\label{fig:50e_ExpRes_Pinata} Experimental results (Averaged $ge$ and Box Plot of final $ge$ values) on AES\_RA - Piñata [50 Epochs]}
\end{figure}

\begin{figure}[!htb]
	\centering
    \subfloat{\includegraphics[width=0.49\textwidth]{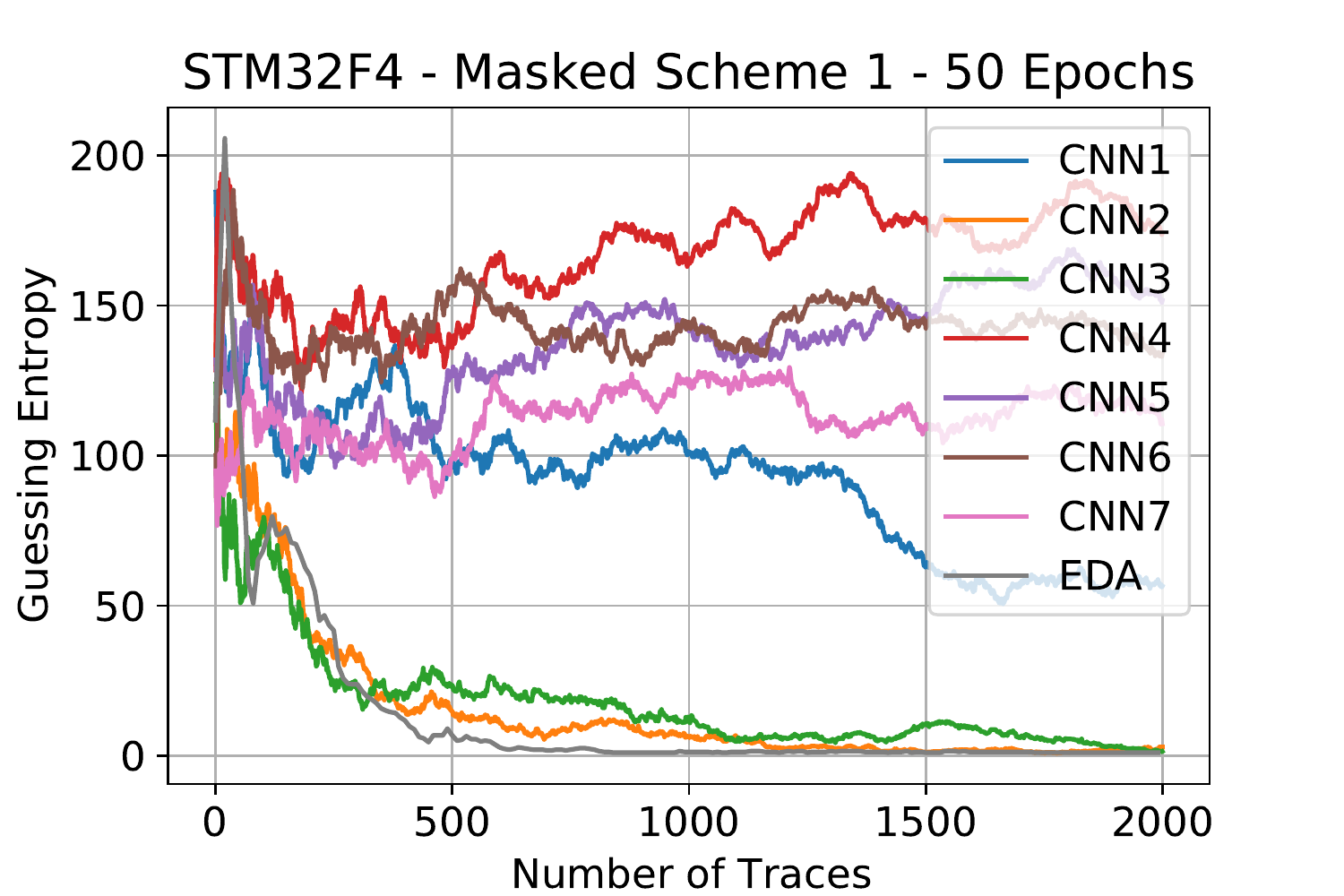}} 
    \subfloat{\includegraphics[width=0.49\textwidth]{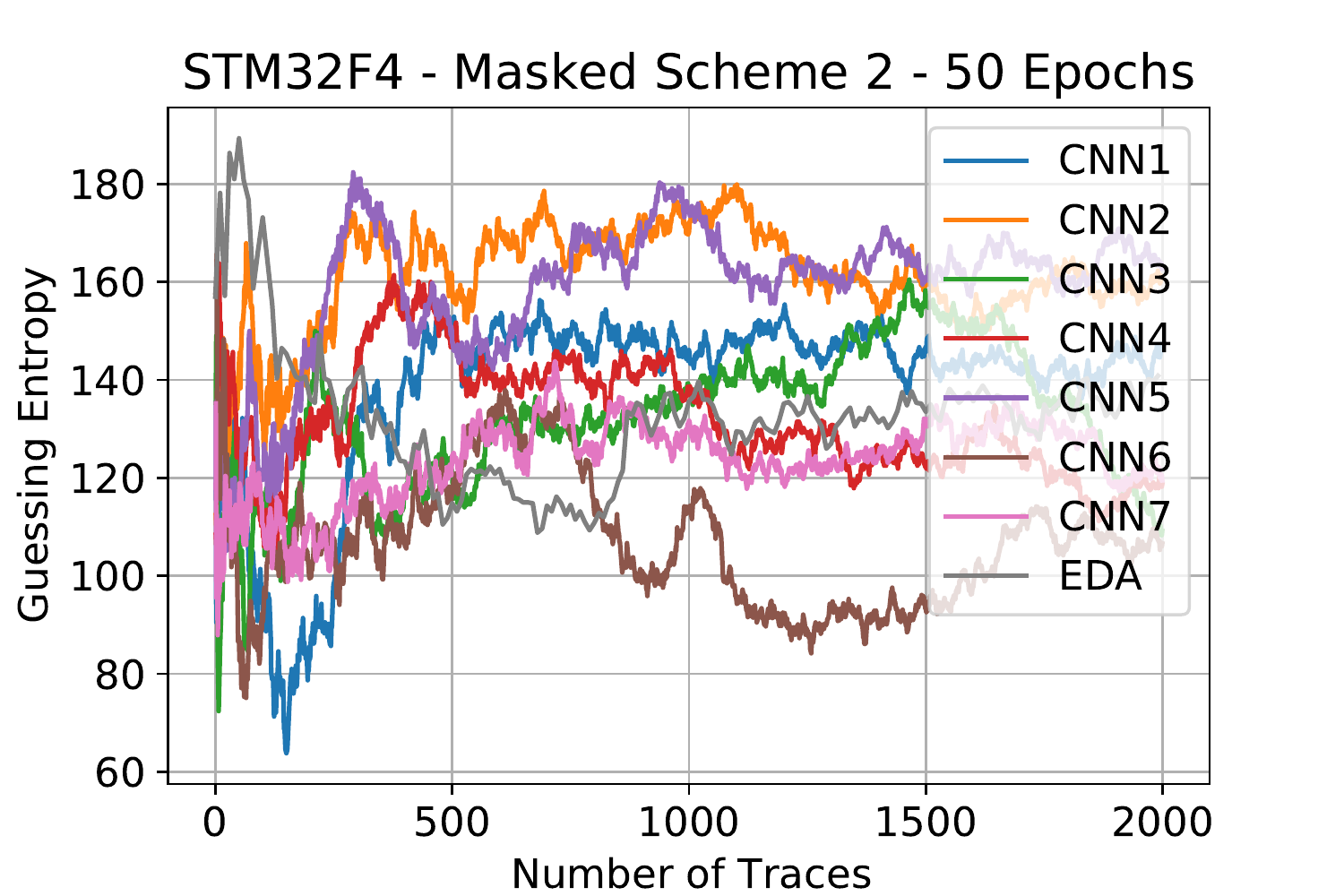}}\\
    \subfloat{\includegraphics[width=0.49\textwidth]{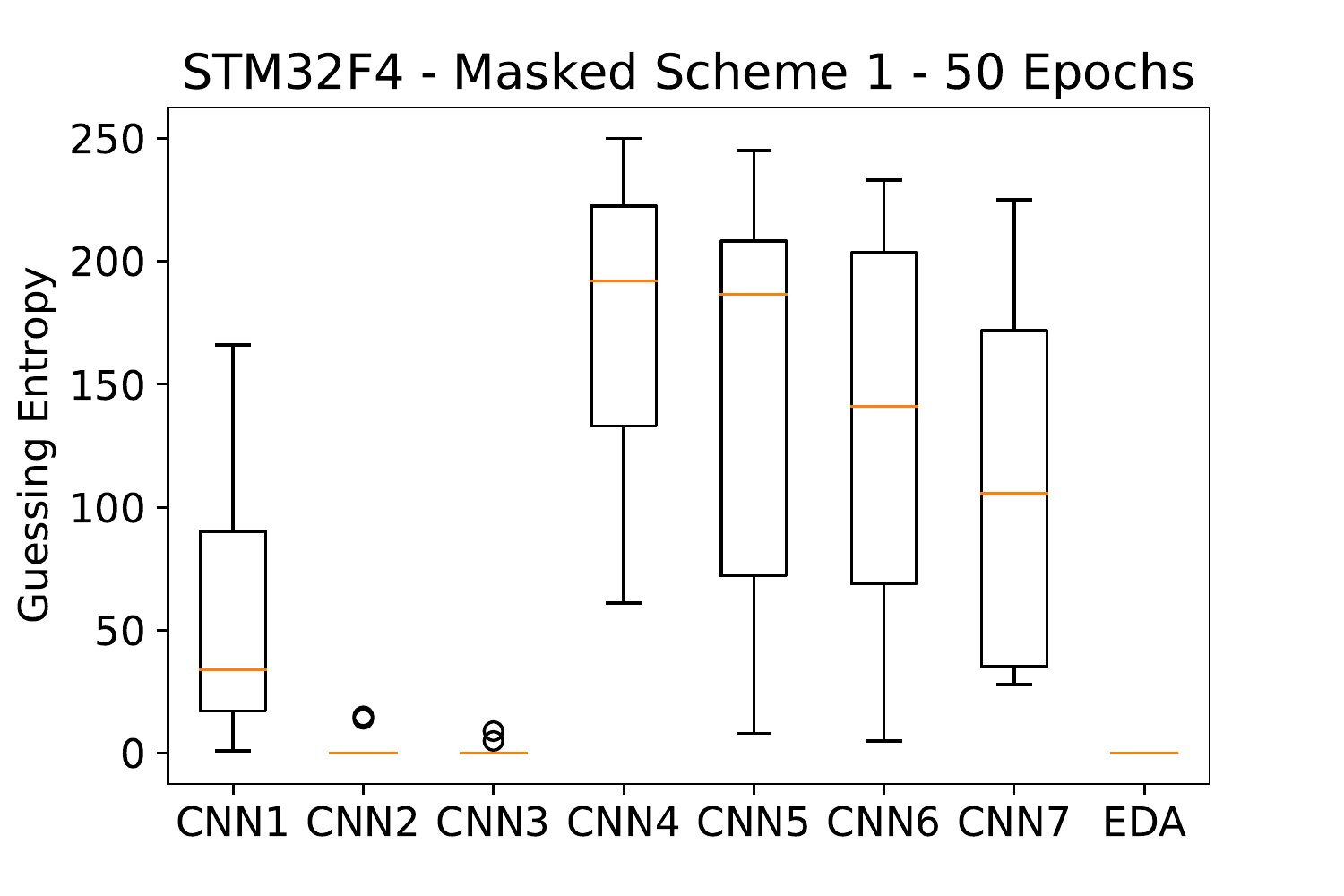}} 
    \subfloat{\includegraphics[width=0.49\textwidth]{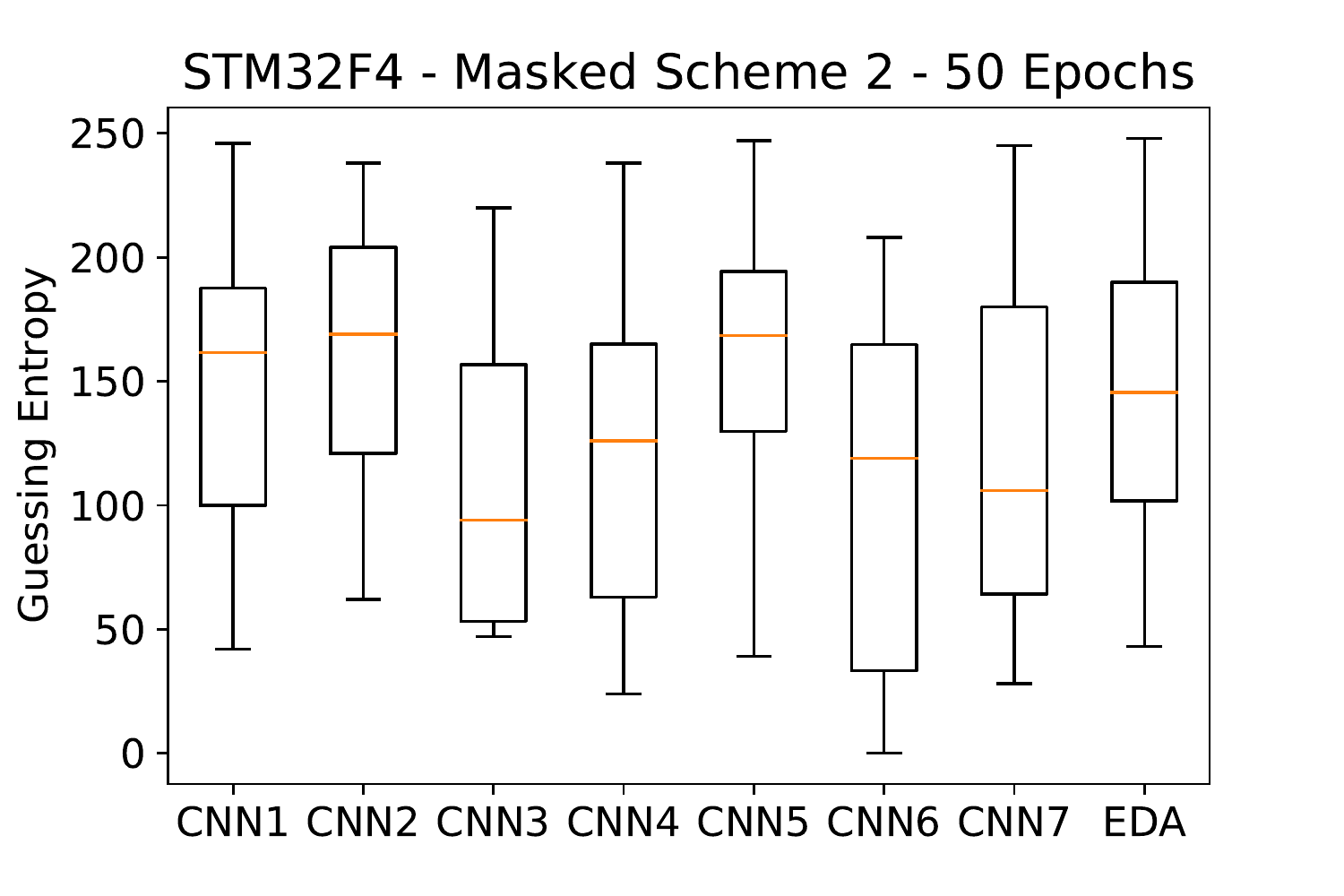}} 
    \caption{\label{fig:50e_ExpRes_STM32F4} Experimental results (Averaged $ge$ and Box Plot of final $ge$ values) on AES\_RA - STM32F4 [50 Epochs]}
\end{figure}

\begin{figure}[H]
	\centering
    \subfloat{\includegraphics[width=0.49\textwidth]{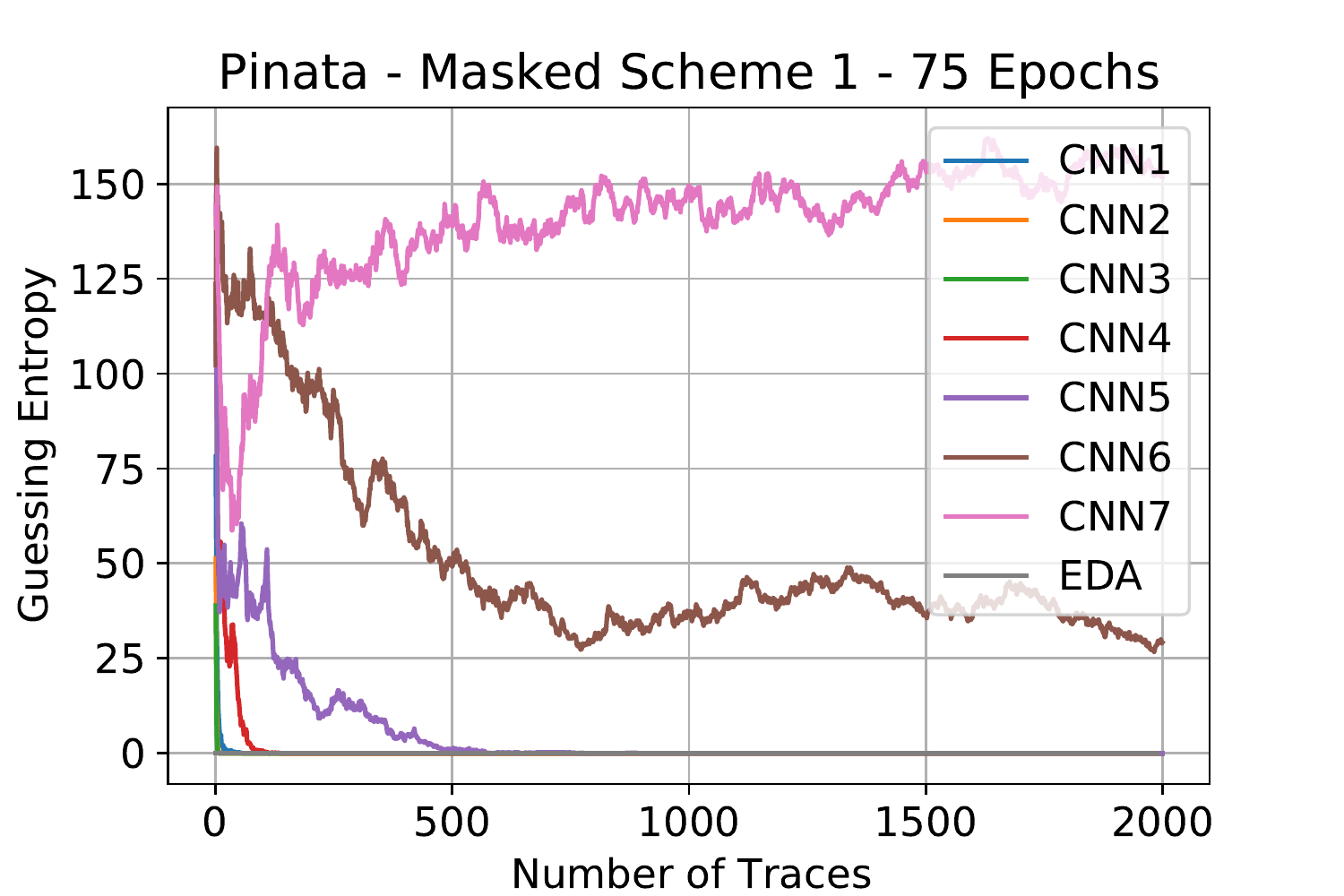}} 
    \subfloat{\includegraphics[width=0.49\textwidth]{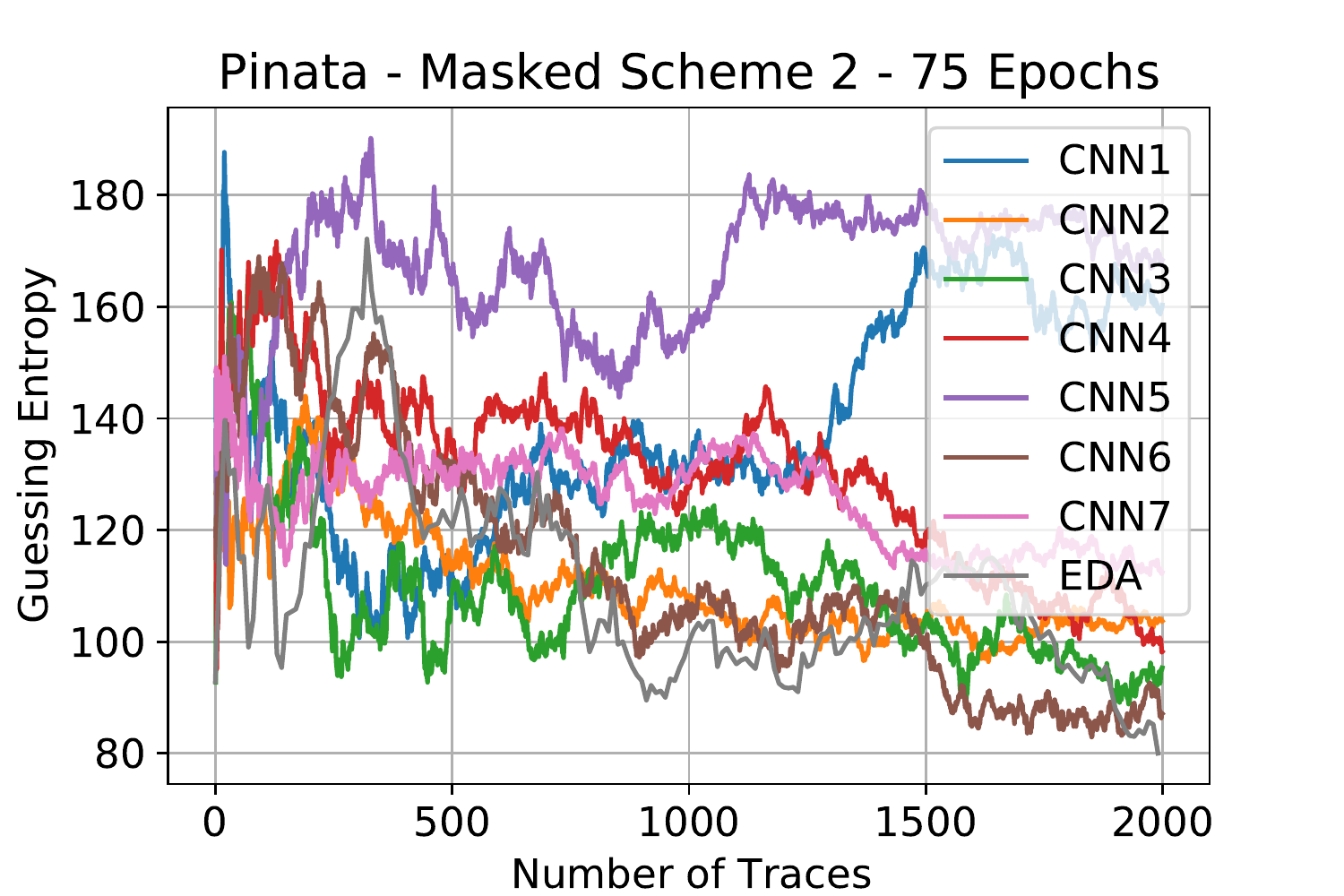}}\\
    \subfloat{\includegraphics[width=0.49\textwidth]{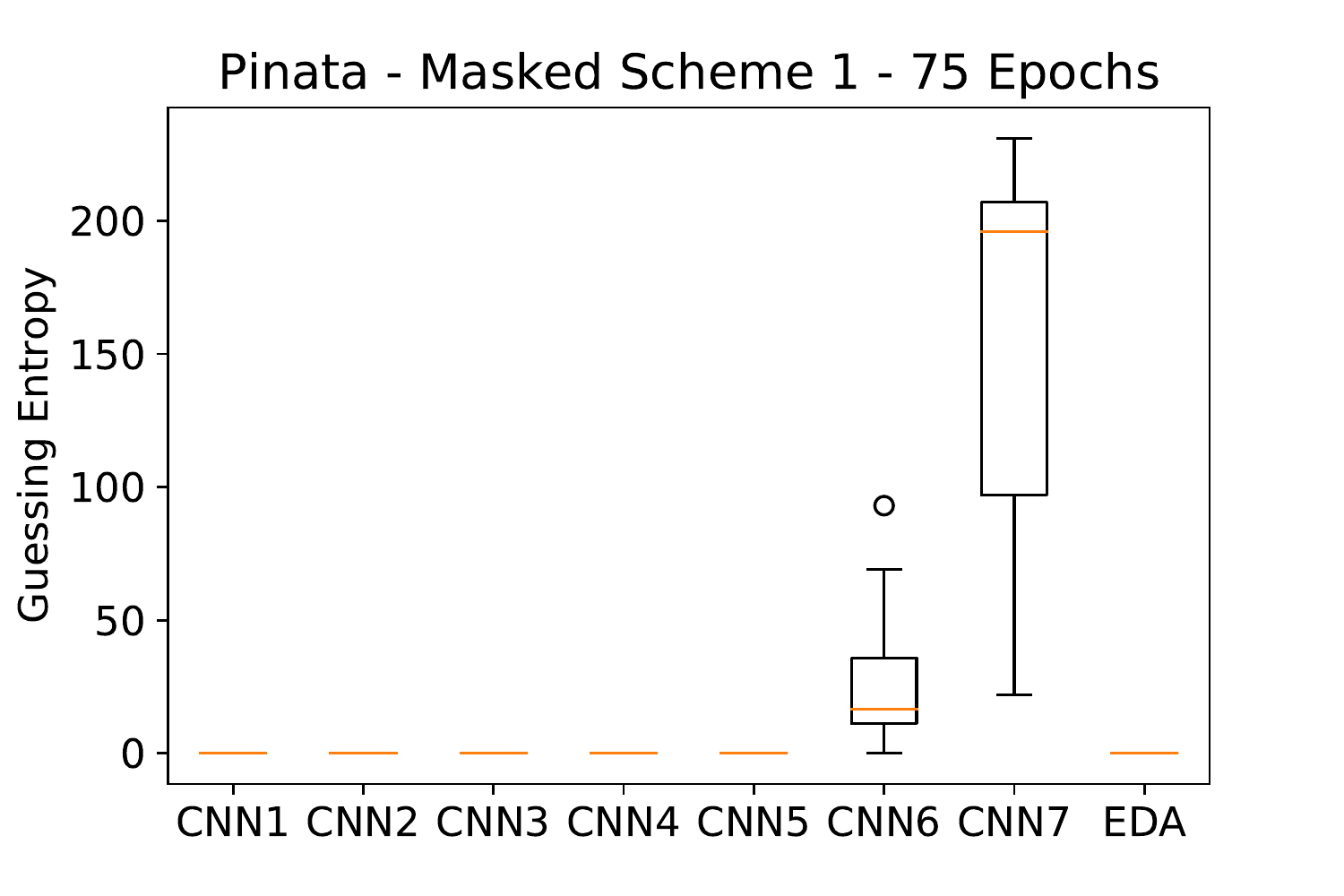}} 
    \subfloat{\includegraphics[width=0.49\textwidth]{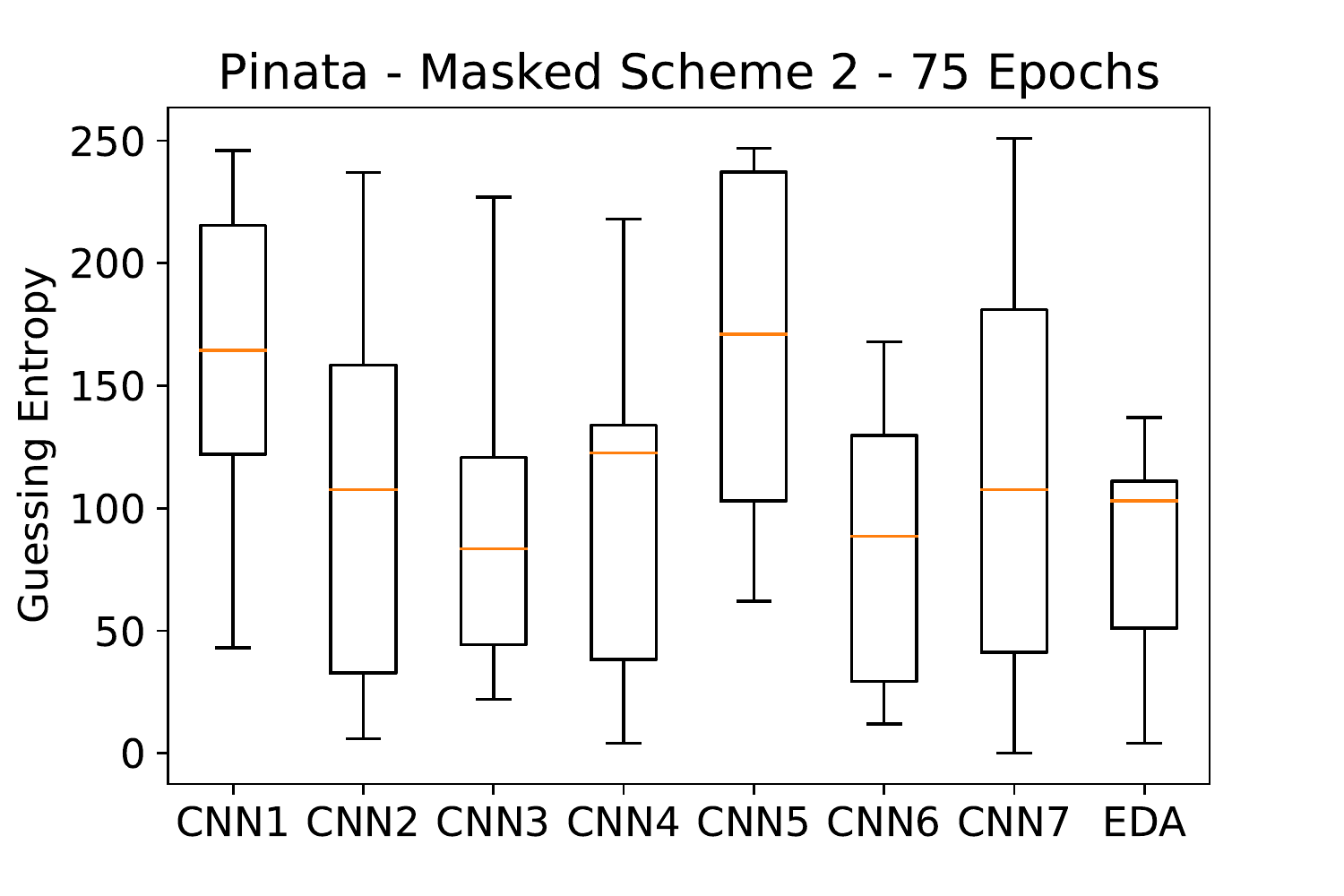}} 
    \caption{\label{fig:75e_ExpRes_Pinata} Experimental results (Averaged $ge$ and Box Plot of final $ge$ values) on AES\_RA - Piñata [75 Epochs]}
\end{figure}

\begin{figure}[!htb]
	\centering
    \subfloat{\includegraphics[width=0.49\textwidth]{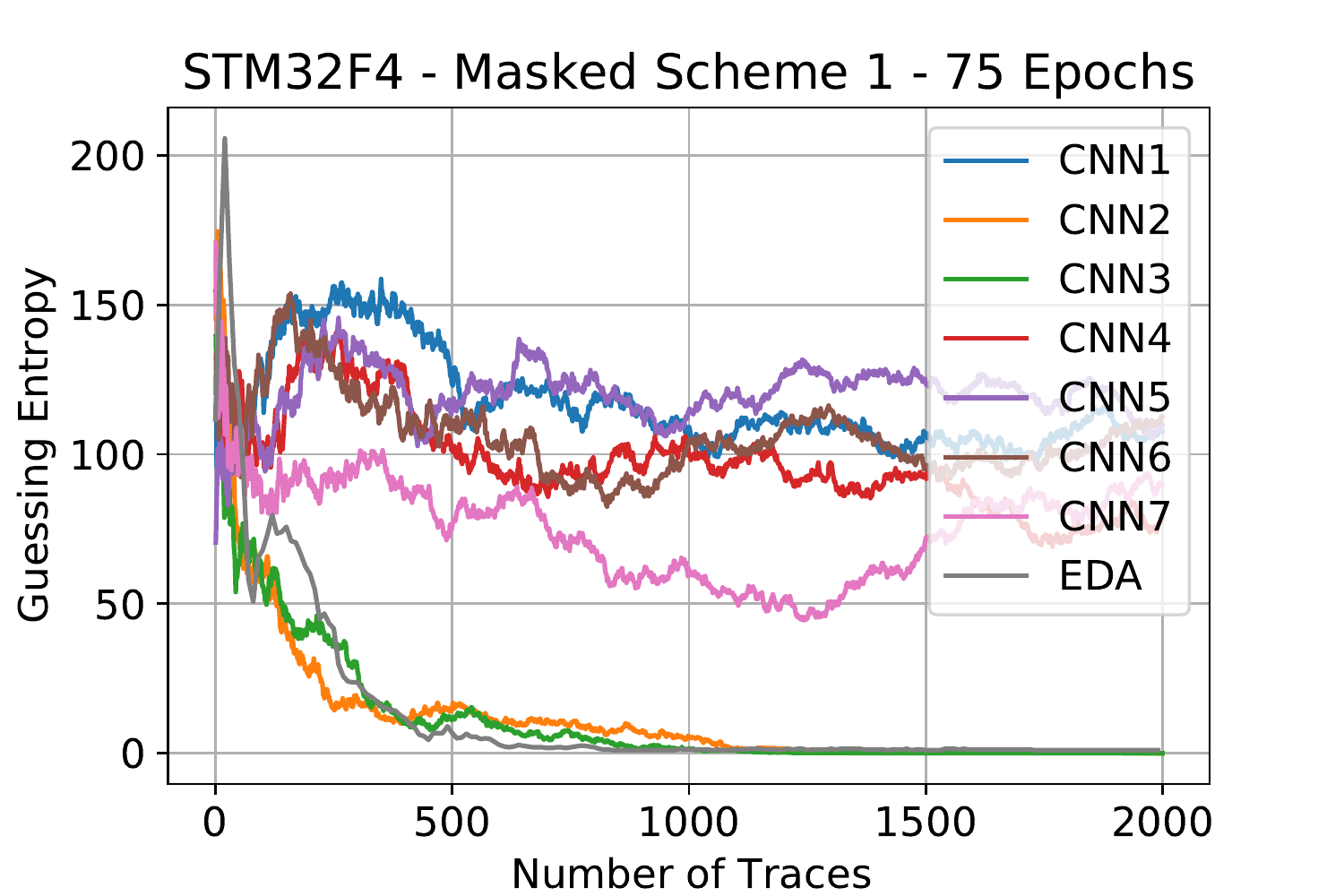}} 
    \subfloat{\includegraphics[width=0.49\textwidth]{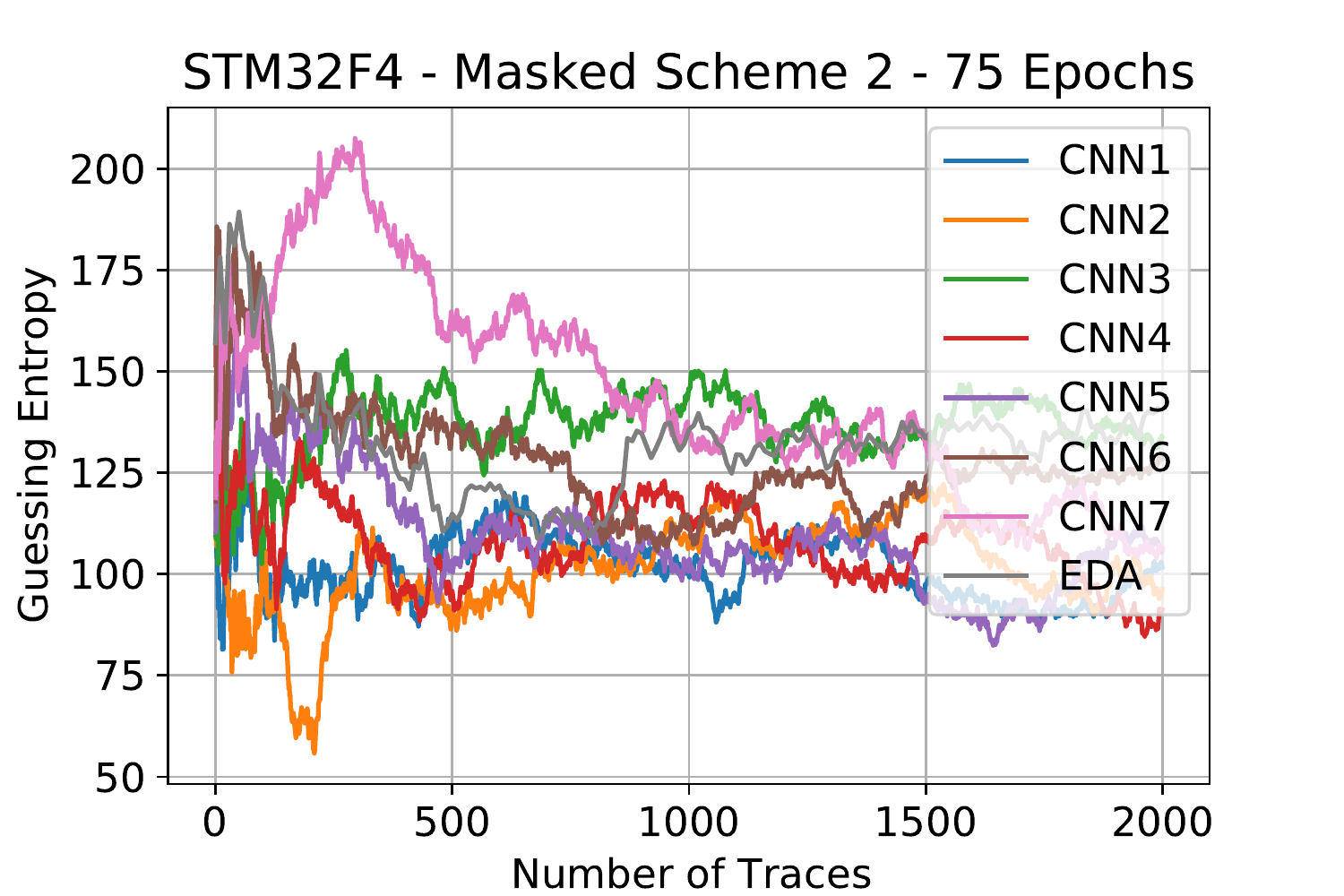}}\\
    \subfloat{\includegraphics[width=0.49\textwidth]{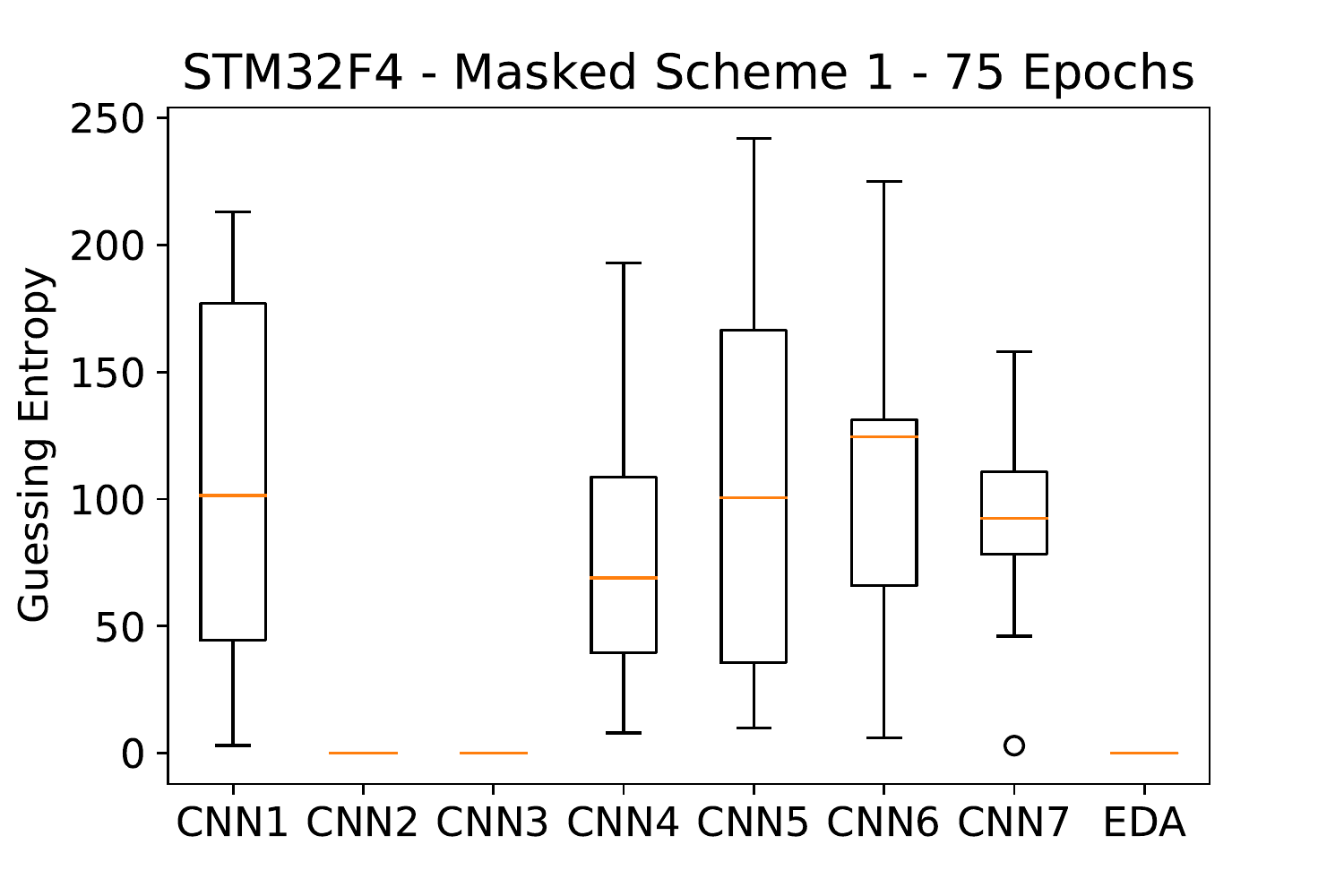}} 
    \subfloat{\includegraphics[width=0.49\textwidth]{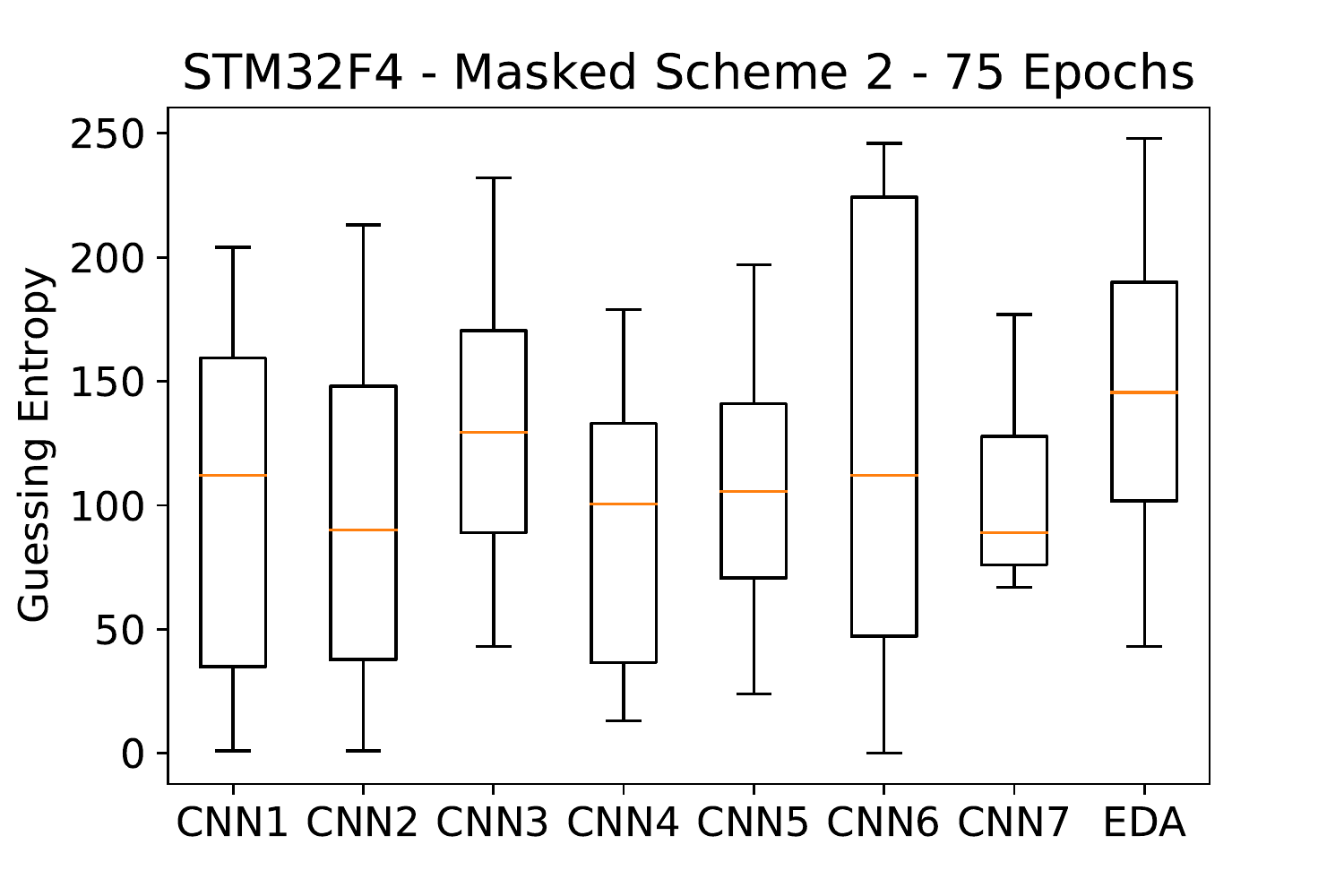}} 
    \caption{\label{fig:75e_ExpRes_STM32F4} Experimental results (Averaged $ge$ and Box Plot of final $ge$ values) on AES\_RA - STM32F4 [75 Epochs]}
\end{figure}

Figures~\ref{fig:50e_ExpRes_Pinata},~\ref{fig:50e_ExpRes_STM32F4},~\ref{fig:75e_ExpRes_Pinata} and~\ref{fig:75e_ExpRes_STM32F4} show the averaged $ge$ of the correct key byte for all CNNs and setups (for 50 and 75 epochs respectively), and also the results using EDA-based PAs. This averaged $ge$ represents the average of 10 attacks using the same model, cross-validated to avoid bias. Also, Table~\ref{tab:ParametersCNNs} shows the number of trainable parameters of each architecture, and the $\bar{Q}_{t_{ge}}$ of successful attacks. Note that although MS1 is especially weak on Piñata, some neural networks could not disclose the key.

\begin{table}[!h]
    \centering
    \resizebox{1\textwidth}{!}{
    \begin{tabular}{|c|C|C|C|C|C|C|C|C|C|}
        \hline
        \textbf{AES\_RA} & \textbf{CNN1} & \textbf{CNN2} & \textbf{CNN3} & \textbf{CNN4} & \textbf{CNN5} & \textbf{CNN6} & \textbf{CNN7} & \textbf{EDA} \\
        \hline
        \textbf{Trainable Param.} & 70 846 848 & 32 960 & 33 070 & 78 172 & 3 418 &  3 471 & \textbf{1 394} & \textbf{1 500} \\
        \hline
        $\mathbf{\bar{Q}_{t_{ge}}}$(\textbf{Piñata-MS1}) & 27 & 8 & 6 & 113 & 34 & - & - & \textbf{5} \\
        \hline
        $\mathbf{\bar{Q}_{t_{ge}}}$(\textbf{STM32-MS1}) & - & 1 867 & 1 222 & - & - & - & - & \textbf{850} \\
        \hline
    \end{tabular}}
    \caption{Top results with PAs on AES\_RA}
    \label{tab:ParametersCNNs}
\end{table}

In addition, the results with DL-based SCA are more variable in general. To show this, Figures~\ref{fig:50e_ExpRes_Pinata},~\ref{fig:50e_ExpRes_STM32F4},~\ref{fig:75e_ExpRes_Pinata} and~\ref{fig:75e_ExpRes_STM32F4} also include Box Plots. These plots represent the variation of the final $ge$ values of each one of the 10 attacks (for each model). In them, the dispersion and the median of the final values can be easily identified, besides clearly distinguishing whether there are outliers. This helps us to determine how precise is the model.



\section{Discussion}
\label{sec:Discussion}

In terms of $ge$, similar outcomes can be achieved with both methods. Nevertheless, it should be noticed that we have only been able to achieve the results reached with EDAs with some CNNs. In addition, as the Box Plots show, the outcomes are in general more variable when using CNNs in our experiments. Also note that although these CNNs were designed for a similar dataset, they usually require human engineering to succeed.

Our comparison shows that EDA-based attacks have some advantages over the DL approach:

\begin{itemize}
    \item Simplicity: fewer parameters to tune (i.e., hyperparameters), trainable parameters, and lighter algorithmic complexity.
    \item Generalized solution: no need to re-tune the hyper-parameters when targeting new devices, implementations, datasets, etc.
    \item Greater control: we can give information about which time samples are expected to be relevant (i.e., which ones have a stronger correlation with the intermediate values).
    \item Search (i.e., “Training”) based on the outcome of realistic key recovery attacks, and not on minimizing a loss function.
    \item The expressiveness and transparency of the probabilistic model that guides the search~\cite{Rioja2021AutoTune}.
\end{itemize}
On the other hand, we have also seen how a well trained DL model can usually approach, or even beat, EDA-based TAs. In addition, training a neural network today is a highly optimized process that can be performed relatively quickly (thanks to the use of parallelization using GPUs) while EDA-based attacks have a long way to go to reach these levels of optimization. However, EDA-based PAs are far less complex than DL in terms of algorithmic complexity. This, together with the fact that several ways of optimizing them have already been identified~\cite{Rioja2021AutoTune} (e.g., attack parallelization, attack computation optimization, etc.), means that they could become as efficient or even more efficient than DL. Conversely, TAs work with a Gaussian assumption, whereas ML models do not assume the probability density function of the data~\cite{Maghrebi2020Deep}. Nevertheless, note that ML models could also be employed with the EDA approach, although authors chose TA for demonstrating their approach in~\cite{Rioja2021AutoTune}.

\section{Conclusions and Future work}
\label{sec:Concl}

From this analysis, we draw the same conclusions about MS1 and MS2 as in~\cite{Rioja2021Elimination}: we recovered the secret key on MS1 using several predefined CNN architectures but not on MS2. This shows that both TAs and DL can circumvent masking in schemes like ASCAD or MS1, as some previous works have also shown~\cite{kim2018noise,prouff2018ascad,Zotkin2018DeepLV}. However, determining whether more complex CNNs or (EDA-based) PAs can actually bypass masking under unfavourable conditions (no mask leakage in the attacked window and/or no unintended interactions) is beyond the scope of this paper, we believe it is an interesting research question for future work.

Concluding, as we intend to show in this paper, both alternatives can provide similar results, with EDA-based attacks being a simpler and more straightforward alternative that can represent a very efficient and interpretable shortcut for evaluators.

\bibliographystyle{splncs04}
\bibliography{references}

\end{document}